\documentclass[preprint,prb,aps,draft]{revtex4}
\begin{document}
\title{Adsorption and Diffusion of Pt and Au on the 
Stoichiometric and Reduced TiO$_2$ Rutile
(110) Surfaces}
\author{Hakim Iddir and Serdar \"{O}\u{g}\"{u}t}
\affiliation{Department of Physics, University of Illinois at Chicago,
Chicago IL 60607}
\author{Nigel D. Browning}
\affiliation{Department of Chemical Engineering and Materials Science, 
University of California-Davis, Davis, CA 95616, and NCEM, Lawrence Berkeley 
National Laboratory, Berkeley, CA 94720}
\author{Mark M. Disko}
\affiliation{Corporate Strategic Research, ExxonMobil Research and
Engineering, Annandale, NJ 08801}
\date{\today}
\begin{abstract}

A comparative first principles pseudopotential study of
the adsorption and migration profiles of 
single Pt and Au atoms on the 
stoichiometric and reduced TiO$_2$ rutile (110) surfaces is presented.
Pt and Au behave similarly with respect to (i) most favorable adsorption 
sites, which are found to be the hollow and substitutional sites on the 
stoichiometric and reduced surfaces, respectively, 
(ii) the large increase in
their binding energy (by $\sim$ 1.7 eV) when the surface is reduced, 
and (iii) their low migration barrier near 0.15 eV on the stoichiometric surface.
Pt, on the other hand, binds more strongly (by $\sim$ 2 eV) to both surfaces.
On the stoichiometric surface, Pt migration pattern
is expected to be one-dimensional, which is primarily influenced by 
interactions with O atoms. Au migration is expected 
to be two-dimensional, with Au-Ti interactions playing a more important role.
On the reduced surface, the migration barrier for Pt diffusion is significantly 
larger compared to Au.

\end{abstract}  
\maketitle

TiO$_2$ is a wide spread catalyst and catalyst support. Because of
its fundamental and technological importance, it has been the
subject of many experimental and theoretical
studies.\cite{Diebold} Pt/TiO$_2$ and Au/TiO$_2$ are two of
the most active catalysts for CO oxidation reactions. In extended
surfaces, Au is not active, but it turns into a very active
catalyst for CO oxidation at low temperatures when highly
dispersed on TiO$_2$.\cite{Haruta,Valden} Au/TiO$_2$ which is
more sensitive to the preparation method than Pt/TiO$_2$, presents
in some cases a higher activity for heterogeneous CO oxidation.
\cite{Bamwenda} Another fundamentally and technologically
important phenomenon which involves the interaction occurring
between small metallic particles and TiO$_2$ surfaces is that of
the strong-metal-support-interaction (SMSI).\cite{Tauster} The
SMSI can induce drastic changes in the performance of the
catalyst, such as suppression of CO and H$_2$ chemisorption, and
has significant effects on structure sensitive reactions.
Pt/TiO$_2$ has been the prototype system for SMSI, while
Au/TiO$_2$ does not undergo significant encapsulation under
equivalent annealing conditions.\cite{Zhang} A fundamental comparative study
of the interaction of these two precious metal catalysts with both
stoichiometric and oxygen deficient TiO$_2$ surfaces will, therefore,
contribute to our understanding of the origin of some of the
differences in their catalytic activity as well as the SMSI phenomenon.

Most of the theoretical work on metal-TiO$_2$ surface interactions 
have focused on the Au/TiO$_2$ interface.\cite{Lopez,Wahlstrom,Vittadini,
Yang,Wang,Giordano,Thien-Nga,Okazaki,Vijay,Mavrikakis}
These studies have covered the structural and electronic 
structure properties and their relationships to the catalytic properties of
the catalyst. On the other hand, in spite of being the prototype SMSI
system, only three self-consistent-field calculations have been performed
for a single Pt atom on TiO$_2$, 
which are also limited due to small system size
and absence of full atomic relaxations.\cite{Thien-Nga,Horsley,We-xing}
Furthermore, all studies on Au, Pt/TiO$_2$ have
considered the adsorption of single or a few atom metal 
clusters on some specific sites of the TiO$_2$ surface.
While these studies have provided useful information on the structural
energetics of metal/support system, a microscopic understanding 
of the interaction occurring between the metal particles and the TiO$_2$
surface also requires calculations for their surface diffusion profiles 
on the stoichiometric and oxygen deficient surfaces. The knowledge of the
surface diffusion profiles is important to
understand the growth mode and evolution of the metal particles
on these surfaces, as the catalytic activity of Au, for example, 
was shown to be particularly structure sensitive.\cite{Valden,Bamwenda,
Mavrikakis,Xu}
The significant differences
in the energy barriers between Pt and Au could also help achieve
a microscopic understanding for the 
occurrence of SMSI in the case of Pt but not for Au.
Motivated by these observations, in this work we present the 
first comparative {\em ab initio} calculations for the
adsorption and diffusion of single Pt and Au atoms
on the stoichiometric and reduced TiO$_2$ (110) surfaces. 

Our calculations for single Pt and Au atoms on the TiO$_2$(110) surface, 
were performed using the {\em ab initio} pseudopotential
total energy method in a slab geometry.\cite{vasp} We used ultrasoft 
pseudopotentials with a cutoff energy of 300 eV, $1\times2\times2$
Monkhorst-Pack k-point grids, and 
Perdew and Wang parametrization of the generalized gradient 
approximation (GGA).\cite{Perdew} 
We made an extensive study of the effects of spin 
polarization effects and found them to be negligible
for the Pt/TiO$_2$ system. For the Au/TiO$_2$ system, including
spin polarization reduced the magnitude of the binding energies
at different selected sites by a maximum amount of $\sim$ 0.13 eV. 
However, the changes in the relative energies between different sites
(which are relevant for the potential energy surfaces) were negligible
(within 0.05 eV). Hence, the calculations for migration energy 
profiles were performed without spin polarization.
After a
systematic study and comparison with larger symmetric slabs, 
we found a 4-layer asymmetric slab, in which 
the two bottom layers are kept at bulk positions, to be the smallest good 
presentation of the surface with and without a metal atom on the surface.
We used a vacuum region of 12 \AA.
To model the single metal 
atom on the surface, 
Pt,Au/TiO$_2$ calculations were performed with $2\times 1$ (stoichiometric)
and $3\times 1$ (reduced) surface unit cells corresponding to 48 and 72 atom
slabs, respectively.
The lattice parameters calculated within GGA ($a=4.641$ \AA, $c=2.982$ \AA, 
and $u=0.305$) are in good agreement with the experimental values
($a_{\rm exp}=4.594$ \AA, $c_{\rm exp}=2.985$ \AA, and $u_{\rm exp}=0.305$)
as well as other theoretical calculations.\cite{lattice}
For the bare TiO$_2$ surface, we find a surface energy of 0.60 J/m$^2$ with
a 7-layer slab, also in good agreement with other 
calculations.\cite{Bates}

We first investigate the binding of a metal atom to the stoichiometric
TiO$_2$ rutile (110) surface. The various possible sites of adsorption are 
shown in Fig. 1(a). We find the most 
favorable position for Pt adsorption on the stoichiometric surface
as the hollow site H1 [Fig. 1(b)]. 
We define the adsorption energy $E_{\rm ads}(M)$
(the gain in energy when a metal atom $M$ binds to a specific site) in terms 
of the total energy of the metal $M$ on the TiO$_2$ surface
$E(M/{\rm TiO}_2)$, the energy of the free surface $E({\rm TiO}_2)$, 
and the energy of a free metal atom $E_{M,{\rm free}}$ as 
$E_{\rm ads}=E({\rm TiO}_2)+E_{M,{\rm free}}-E(M/{\rm TiO}_2)$.
The calculated adsorption energy of a Pt atom on the hollow site is 
$E_{\rm ads}{\rm (Pt)}=2.51$ eV. The least favorable position for Pt
adsorption, by $\sim$ 1 eV compared to H1, is the Ti5c site. This 
finding is in sharp contrast with an interpretation from an experimental 
study\cite{Schierbaum} which suggests that Pt atoms at low coverages
adsorb preferentially at the Ti5c site. 
However, only two positions, which were based on the
key assumptions of hard-sphere atoms with van der Waals type
interactions, were considered as possibilities in the experimental work.
Furthermore, the ratio of the attenuated core level
photoemission intensities of the Ti5c atom and in-plane surface O atom, 
which was used to assign the preferential adsorption on the Ti5c
site, is {\em not} inconsistent with our finding of the hollow
site the most favorable site. This is because the Pt
atom at H1 is slightly tilted with respect to 
its relaxed position directly over the Ti5c site [Fig. 1(b)].
The Pt-Ti5c distance of 2.50 \AA\ (at H1) 
is only 0.12 \AA\ larger than the relaxed Pt-Ti5c distance when it sits
directly on top of Ti5c. What makes the hollow site the most 
energetically favorable position is predominantly the formation 
of a Pt-O bond at this site. An examination of Table I
shows that Pt-O bonding plays an important
role in the structural energetics of Pt on the surface, since 
the lower energy sites are those in which 
Pt and oxygen are nearest neighbors, and the formation 
of a Pt-Ti bond is a secondary effect. In fact, in the presence
of Pt on the surface, the distance between the bridging oxygen O1
and the six-fold coordinated Ti (Ti6c) increases to 1.98 \AA\
from its bare surface value of 1.84 \AA, as Pt pulls the
bridging oxygen significantly toward it at the expense of increasing 
the surface Ti6c-O1 distance. 

The most favorable site for Au adsorption on the 
stoichiometric surface is also the hollow site. However, the calculated
adsorption energy $E_{\rm ads}({\rm Au})=0.58$ eV is lower by almost 2 eV
compared to Pt. This result is in very good agreement with previous 
calculations,\cite{Wang,Vijay} which find adsorption energies near 0.6 eV.
The lower adsorption energy also correlates with the 
considerably larger distances of Au to O1 and Ti5c atoms, which are 
calculated to be 2.38 \AA\ and 2.85 \AA, respectively. As such, the 
surface Ti6c-O1 distance, in the presence of Au, is 1.89 \AA, only 
slightly larger than the bare surface value. This means that, 
in comparison to Pt, Au atoms do not perturb the
stoichiometric TiO$_2$ (110) surface significantly.
The relative stability of various sites for Au adsoprtion is also different
compared to Pt. In particular, Ti5c site (the least favorable site for Pt
adsorption) is only 0.1 eV higher in energy compared to the hollow site, and 
Au-O bonding plays only a minor role compared to Au-Ti bonding. 

In order to achieve a better understanding of the similarities and the 
differences between Pt and Au adsorption and to gain insight into the
energetics of their diffusion, we calculated the 
full migration energy profiles of the two metals. On the stoichiometric
surface, a rectangular region of dimensions $c/2$ (along $[001]$) and
$a/\sqrt{2}$ (along $[1\overline{1}0]$) was divided into a $4\times 5$
uniform grid. Due to the reflection symmetries, the energy profile will 
be symmetric with respect to the chosen rectangular region. The metal atom 
was placed at each grid point and allowed to relax only in the direction
perpendicular to the surface. All the Ti and O atoms in the top two layers
were allowed to relax as well. 
The results interpolated to a finer $(30\times 30)$ grid 
are displayed in Figures 2(a) and 2(b) 
for Pt and Au, respectively. For Pt, the
energy surface is quite corrugated with a large variation of $\sim$ 1 eV
across the entire grid. The minimum and maximum occur at 
the hollow site and the Ti5c positions, in agreement with the results 
displayed in Table I. For a Pt atom at the hollow site, barrier to 
migration along the $[1\overline{1}0]$ direction is 0.95 eV.
Such a large barrier is expected to 
make Pt migration in the $[1\overline{1}0]$ direction unlikely, except
at high temperatures. In the perpendicular $[001]$ direction, 
on the other hand, Pt diffusion from one hollow site to the next has a
very small barrier as the Pt atom diffuses over the basal plane
oxygen. The barrier for this migration is 0.13 eV. The importance of 
a Pt-O bond is evident, as the Pt atom can diffuse between hollow
sites along the $[001]$ direction easily by bonding to bridging oxygen
(at the hollow site) and to the basal oxygen at favorable Pt-O distances
near 2.01 \AA\, as observed in PtO and Pt$_3$O$_4$ compounds.
This theoretical finding of an easy channel for Pt diffusion is in good 
agreement with an STM study of 
Pt/TiO$_2$,\cite{Dulub} which suggests that 
$[001]$ is the preferred orientation for Pt diffusion, based on the 
elongated shapes of the Pt clusters on TiO$_2$ (110).

In contrast to Pt, the potential energy profile for Au migration [Fig. 2(b)]
is quite flat, with energy barriers along any direction not exceeding 0.35 eV.
The minimum and maximum occur at the hollow site and between the bridging
oxygens (above Ti6c), respectively. The barriers for Au migration along
$[1\overline{1}0]$ and $[001]$ are 0.17 eV and 0.2 eV, respectively.
Such low barriers along both directions 
indicate that Au can diffuse rather easily on 
the stoichiometric surface, and could also explain the wide variation 
in the minimum energy positions of Au on TiO$_2$(110) reported
in the literature.\cite{Lopez} 
The theoretical values obtained here are in good agreement with
the experimental observation of easy Au diffusion on the oxide
surface even at 140 K, as well as the estimates
for the Au adsorption energy of 0.5 eV and its small migration barrier 
of 0.07 eV.\cite{Parker}
The flat surface profile is also consistent with the rapid increase in Au
cluster size observed at very low Au coverages and the change 
in the Au particle size distribution to bimodal, upon oxidizing
the reduced surface (Ostwald ripening), which is favored by fast 
surface diffusion of atomic and small Au clusters.\cite{Lai}

The migration of a single Au atom 
on the TiO$_2$(110) surface is expected to be two-dimensional. Pt migration,
on the other hand, is strongly one-dimensional, along the $[001]$ easy 
channels, due to the large barrier near 1 eV along the
$[1\overline{1}0]$ direction. 
In spite of the significant differences between 
the migration profiles of the two metals, they both have a small
minimum energy barrier near 0.15 eV for their migration on the 
stoichiometric surface. For Pt migration, the easy channel involves
a somewhat continuous bonding of the Pt atom to the surface (bridging
and basal) oxygens. This suggests that 
the structural energetics of Pt on the TiO$_2$(110) surface is primarily
influenced by the interaction of the $p$ bands of the oxygen and 
the $d$ orbitals of Pt. 
Since Au has a complete $d-$shell, the $p-d$ hybridization is not 
expected to play a significant role. The first principles calculations
presented here indeed support this picture: The Au-Ti distance at the 
hollow site is 2.85 \AA, the same as the nearest neighbor distance
in the compound AuTi. At the Ti5c position, the Au-Ti distance is 
2.78 \AA, only slightly shorter than this value. However, as Au gets
farther away from Ti, and closer to surface oxygen, the potential 
energy profile indicates higher (albeit still less than 0.35 eV) barriers. 
These observations suggest that the structural energetics of a gold atom 
is primarily influenced by the interaction of the Au $s$ orbitals with 
the Ti $d$ band. 

We now move to the reduced surface, where one of the bridging oxygens is 
missing. For both Pt and Au, the most favorable adsoprtion site is the 
substitutional one (the site of the oxygen vacancy). The adsorption energies
$E_{\rm ads}$ at this site are significantly larger compared to 
the stoichiometric surface, 4.28 eV (Pt) and 2.18 eV (Au).
The distances of the metal atom to the 
nearest neighbor Ti6c site are 2.39 \AA, and 2.67 \AA\ for Pt and Au, 
respectively. As in the stoichiometric surface, Pt sits closer to the 
surface and distorts the underlying lattice more significantly. 
We calculated the full migration barrier profile
of the two metal atoms on the reduced surface. For this case, we chose a 
rectangular region of dimensions $c$ (along $[001]$) and $a/\sqrt{2}$
(along $[1\overline{1}0]$) with the substitutional site at one of the 
corners. The results are shown
in Figures 3(a) for Pt and 3(b) for Au. Clearly, both metals
show a strong preference for adsorption on the substitutional site
with similar potential energy profiles both along the $[001]$ 
and $[1\overline{1}0]$ directions. The migration barrier for both 
metals along the $[001]$ direction (toward the bridging oxygen)
is rather high, 1.85 eV for Au and 2.3 eV for Pt. This is in contrast
to the stoichiometric surface where the $[001]$ direction has a small
energy barrier along the neighboring hollow sites. The smallest 
migration barrier on the reduced surface occurs along the 
$[1\overline{1}0]$ direction, from the substitutional site toward
the Ti5c position. In fact, this is where the main difference
between Au and Pt is observed. Although the migration barrier for 
Pt along this direction is still high (1.1 eV), the barrier for 
Au is only 0.6 eV. This implies that while oxygen vacancies act
as anchoring sites for metal adsorption on the reduced surface, 
Au is much more likely to escape from the relatively shallower
potential well created by the vacancy compared to Pt, which 
gets trapped at this site. In addition, Figure 3(b) shows that
in the case of Au, there exists an extra minimum in the potential 
energy profile near the diagonal Ti5c site. The energy of this
site is only 0.3 eV higher than that of the substitutional site.
The existence of this extra minimum at about the same
location as the hollow site of the stoichiometric surface shows
that the effect of the vacancy for the Au/TiO$_2$ system is more
localized than that for the Pt/TiO$_2$ system.

In conclusion, we have shown that on the stoichiometeric surface, Pt 
displays a one-dimensional migration pattern on a highly 
corrugated potential energy profile, and Au migration is two-dimensional 
with a relatively flat profile. The structural energetics of Pt and Au
are primarily influenced by their interactions with O and Ti, 
respectively. The distortions of the TiO$_2$ support
due to a single atom adsorbate              
are much more pronounced for Pt than Au.
On the reduced surface, while
both Pt and Au are anchored at the substitutional site, Pt binding to 
oxygen vacancies is considerably stronger compared to Au.
From these two observations, we can already infer
even at a single atom level the 
predisposition of Pt/TiO$_2$
rather than Au/TiO$_2$ to undergo encapsulation, which is evidence of SMSI.
These findings could also provide a hint for the higher 
sensitivity of Au catalytic activity to the preparation methods, 
as Au atom binding to the oxide surface is not only weak but also 
rather site-insensitive (flat profile), making its activity 
more vulnerable to external effects.

This work was supported by the ACS Petroleum Research Fund under grant
\#s 40028-AC5M and 37552-AC5, and by NCSA 
under grant \# DMR030053.

\newpage
\begin{figure}[h]
\caption{(a) An oblique view of an extended TiO$_2$ (110) surface showing the 
possible sites for metal adsorption above them (listed in Table I). 
The white and black circles
represent Ti and O atoms, respectively. (b) Top 2-layer view of TiO$_2$ (110)
surface with a Pt atom (gray circle) adsorbed at the hollow site.}
\end{figure}

\begin{figure}[h]
\caption{The potential energy profiles for (a) Pt (b) Au on the 
stoichiometric TiO$_2$ (110) surface. The profiles were doubled 
along the [001] direction (compared to the chosen rectangular 
grid mentioned in the text) to have the same size as the profiles
in the next figure.}
\end{figure}

\begin{figure}[h]
\caption{The potential energy profiles for (a) Pt (b) Au on the 
reduced TiO$_2$ (110) surface}
\end{figure}

\begin{table}
\caption{The relative energies (in eV) for Pt and Au adsorption above the given
specific sites with respect to the most stable (H1) site for
the stoichiometric TiO$_2$ (110) surface.}
\begin{ruledtabular}
\begin{tabular}{lcc}
Adsorption Site & $\Delta E$ (Pt) & $\Delta E$ (Au) \\ \hline
Hollow site (H1) & 0.0 & 0.0 \\
Basal oxygen & 0.11 & 0.27 \\
Hollow site (H2) & 0.13 & 0.20 \\
Between bridging O (Ti6c) & 0.25 & 0.34 \\
Bridging oxygen (O1) & 0.70 & 0.17 \\
5-fold Ti (Ti5c) & 0.95 & 0.10 \\
\end{tabular}
\end{ruledtabular}
\end{table}


\begin{thebibliography}{99}
\bibitem{Diebold} For an excellent review, see
U. Diebold, Surf. Sci. Rep. {\bf 48}, 53 (2003) and references therein.

\bibitem{Haruta} M. Haruta, Catal. Today {\bf 36}, 53 (1997).

\bibitem{Valden} M. Valden, X. Lai, and D.W. Goodman,
Science {\bf 281}, 1647 (1998).

\bibitem{Bamwenda} G. R. Bamwenda, S. Tsubota, T. Nakamura, 
and M. Haruta, Catal. Lett. {\bf 44}, 83 (1997).

\bibitem{Tauster} S. J. Tauster, S. C. Fung, and R. L. Garten, J. Am. Chem. 
Soc. {\bf 100}, 170 (1978).

\bibitem{Zhang} L. Zhang, R. Persaud, and T.E. Madey, 
Phys. Rev. B {\bf 56}, 10549 (1997).

\bibitem{Lindan} P. J. D. Lindan, J. Muscat, S. Bates, 
N. M. Harrison, and M. Gillan, Faraday Discuss. {\bf 106}, 135 (1997).

\bibitem{Lopez} N. Lopez and J. K. Norskov, Surf. Sci. {\bf 515}, 175 (2002).

\bibitem{Wahlstrom} E. Wahlstr\"{o}m, N. Lopez, R. Schaub, P. Thostrup, A.
R\o nnau, C. Africh, E. L\ae gsgaard, J. K. N\o rskov, and F.
Besenbacher, Phys. Rev. Lett. {\bf 90}, 026101 (2003).

\bibitem{Vittadini} A. Vittadini and A. Selloni, 
J. Chem. Phys. {\bf 117}, 353 (2002).

\bibitem{Yang} Z. Yang, R. Wu, and D. W. Goodman, Phys. Rev. B {\bf 61} 14066
(2000).
\bibitem{Wang} Y. Wang and G. S. Hwang, Surf. Sci. {\bf 542}, 72 (2003).

\bibitem{Giordano} L. Giordano, G. Pacchioni, T. Bredow, and J.F. Sanz, 
Surf. Sci. {\bf 471}, 21 (2001).

\bibitem{Thien-Nga} L. Thi\^{e}n-Nga and A. T. Paxton, Phys. Rev. B {\bf 58}, 
13233 (1998).

\bibitem{Okazaki} K. Okazaki, Y. Morikawa, S. Tanaka, K. Tanaka, and 
M. Kohyama, Phys. Rev. B {\bf 69},
235404 (2004).

\bibitem{Vijay} A. Vijay, G. Mills, and H. Metiu, 
J. Chem. Phys. {\bf 118}, 6536 (2003).

\bibitem{Mavrikakis} M. Mavrikakis, P. Stoltze, and J.K. Norskov, 
Catal. Lett. {\bf 64}, 101 (2000).

\bibitem{Horsley} J.A. Horsley, J. Am. Chem. Soc. {\bf 101}, 2870
(1979).

\bibitem{We-xing} X. Wei-xing, K. D. Schierbaum, and W. Goepel, 
J. Sol. State Chem. {\bf 119}, 237 (1995).

\bibitem{Xu} Y. Xu and M. Mavrikakis, J. Phys. Chem. B {\bf 107}, 9298 (2003).

\bibitem{vasp} G. Kresse and J. Hafner, Phys. Rev. B {\bf 47}, R558 (1993); 
G. Kresse and J. Furthm\"{u}ller, {\em ibid.} {\bf 54}, 11169 (1996).

\bibitem{Perdew} J.P. Perdew, J.A. Chevary, S.H. Vosko, K.A. Jackson, 
M.R. Pederson, D.J. Singh, and C. Fiolhais, Phys. Rev. B {\bf 46}, 
6671 (1992).
           
\bibitem{lattice}  K. M. Glassford and J. R. Chelikowsky, Phys. Rev. B
{\bf 46}, 1284 (1992); M. Ramamoorthy, R. D. King-Smith, and 
D. Vanderbilt, {\em ibid.} {\bf 49}, 7709 (1994); D. C. Allan and 
M. P. Teter, J. Am. Ceram. Soc. {\bf 73}, 3247 (1990).

\bibitem{Bates}
S. P. Bates, G. Kresse, and M. J. Gillan,
Surf. Sci. {\bf 385}, 386 (1997); 

\bibitem{Schierbaum} K. D. Schierbaum, S. Fischer, M. C. Torquemada, 
J. L. de Segovia, E. Rom\'{a}n, and J. A. Mart\'{\i}n-Gago, Surf. Sci. {\bf 345}, 
261 (1996).

\bibitem{Dulub} O. Dulub, W. Hebenstreit, and U. Diebold, Phys. Rev. Lett.
{\bf 84}, 3646 (2000).

\bibitem{Parker} 
S. C. Parker, A. W. Grant, V. A. Bondzie, and C. T. Campbell, 
Surf. Sci. {\bf 441}, 10 (1999); C. T. Campbell, S. C. Parker, and D. E. Starr,
Science {\bf 298}, 811 (2002).

\bibitem{Lai} X. Lai and W.
Goodman, J. Mol. Catal. A. Chem. {\bf 162} 33, (2000).

\end{thebibliography}
\end{document}